\documentclass[12pt]{article}
\usepackage{amsmath}
\usepackage{graphicx, epsfig}
\usepackage{times}
\input epsf
\begin{document}
\begin{titlepage}
\title{ Energy dependence of  gap survival probability and antishadowing}
\author{ S.M. Troshin\footnote{e-mail: Sergey.Troshin@ihep.ru},  N.E. Tyurin\\[1ex] \small\it Institute
\small\it for High Energy Physics,\\\small\it Protvino, Moscow Region, 142281 Russia}
\date{}
\maketitle
\begin{abstract}
We discuss energy dependence of gap survival probability which
follows from rational form of amplitude unitarization. In contrast
to eikonal form of unitarization which leads to decreasing energy
dependence of gap survival probability, we predict a
non-monotonous form for this dependence.
\end{abstract}
\vfill
\end{titlepage}

Studies of the processes with large rapidity gaps are very important
as a tool for the search of new physics. For the first time such processes
have been discussed in \cite{dok,land,bj}.
Predictive power of  QCD calculations for cross-sections of such
processes is affected by the uncertainties related to the soft interactions
 (rescattering) in initial
and final states.
Dynamics of such interactions is accounted then by the introduction of a factor
which is known as a gap survival probability \cite{bj}, i.e. a probability to keep away
inelastic interactions  which can result in filling up by hadrons
the large rapidity gaps.
Energy dependence and magnitude of  gap
survival probability is an important issue e.g. in the studies
of  Higgs production in double diffractive exclusive and inclusive
processes at Tevatron and the LHC (cf.  \cite{khoze}).  The
extensive studies of magnitude and energy dependence of gap survival probability
have been performed, results of these studies can be found, e.g. in
papers \cite{flet,gots,block,kaid}.

The gap survival probability $\langle |S|^2\rangle$ is determined by the
relation \cite{bj}:
\begin{equation}\label{gap}
\langle |S|^2\rangle=\frac{\int_0^\infty D_H(b)|S(s,b)|^2 d^2b}{\int_0^\infty D_H(b)d^2b},
\end{equation}
where $D_H(b)$ is the probability to observe a specific hard
interaction in collision of the hadrons $h_1$ and $h_2$ and
$P(s,b)\equiv |S(s,b)|^2$, where $S$ is elastic scattering
$S$--matrix, i.e. $P$ is a probability of the absence of the
inelastic interactions. In the eikonal formalism which is usually
used for estimation of $\langle |S|^2\rangle$ the probability
$P(s,b)=\exp(-\Omega(s,b))$. All estimations of the gap
 survival probability performed on the basis of the eikonal amplitude
  unitarization lead to
decreasing energy dependence of this quantity. Therefore  rather small values of
cross--sections  for diffractive Higgs productions are expected
 at the LHC energies \cite{der,petr}.

However, there is an alternative approach to unitarization which
 utilizes a rational representation
and leads, as it will be shown below, to a non-monotonous energy dependence
of gap survival probability.
Arguments based on analytical properties of the scattering
amplitude  \cite{blan}  provide support for  the rational form of unitarization.
In potential scattering
 rational form of unitarization
corresponds to an approximate wave function which changes both
the phase and amplitude of the wave.
The rational form of unitarization in quantum field theory
 is based on the relativistic generalization \cite{umat}
 of the Heitler equation \cite{heit}.
In the $U$--matrix approach based on rational form of
unitarization, the elastic scattering amplitude in the impact
parameter representation has the form:
\begin{equation}
f(s,b)=\frac{U(s,b)}{1-iU(s,b)}, \label{um}
\end{equation}
where $U(s,b)$ is the generalized reaction matrix, which is
considered to be an input dynamical quantity similar to the
eikonal function. Unitarity equation for the elastic amplitude
$f(s,b)$ rewritten at high energies has the form
\begin{equation}
\mbox{Im} f(s,b)=|f(s,b)|^2+\eta(s,b) \label{unt}
\end{equation}
where the inelastic overlap function
\[
\eta(s,b)\equiv\frac{1}{4\pi}\frac{d\sigma_{inel}}{db^2}
\]
 is the sum of
all inelastic channel contributions.
Inelastic overlap function
is related to $U(s,b)$ according to Eqs. (\ref{um}) and (\ref{unt}) as follows
\begin{equation}
\eta(s,b)=\frac{\mbox{Im} U(s,b)}{|1-iU(s,b)|^{2}}\label{uf}.
\end{equation}
The probability
\begin{equation}
P(s,b)\equiv |S(s,b)|^2=\left|\frac{1+iU(s,b)}{1-iU(s,b)}\right|^2\label{prob}.
\end{equation}
Unitarity of the scattering matrix implies, in principle, an
existence at high enough energies $s>s_0$, where $s_0$ is the
threshold,
 of the new scattering mode --
antishadow one. It has been revealed in \cite{bds} and
effects related to antishadowing at the LHC energies have
 been discussed in \cite{lhc}.
The most important feature of this mode is the self-damping of the
contribution from the inelastic channels. The rational form of
unitarization provides smooth transition beyond the black disk
limit, where antishadow scattering mode is realized, i.e. at high
energies and at small impact parameters elastic scattering channel
can  play dominating role.

There is an experimental indication that this mode can indeed
occur at very high energies. The analysis of the experimental data
on high--energy diffractive scattering
 shows that the effective interaction area expands with energy and the
interaction intensity -- opacity -- increases with energy at fixed
impact parameter. At the Tevatron highest energy $\sqrt{s}=1800$
GeV elastic scattering amplitude is very close to black disk limit
at $b=0$ \cite{cdf}, i.e.
\[
\mbox{Im} f(s,b=0)=0.492\pm 0.008.
\]
The central inelastic collisions of hadrons are far from
amalgamation of the two hadrons in one region of space as it was
shown in \cite{chyang} and the persistence of longitudinal
momentum takes place at very high energies.

The rational form of amplitude unitarization can be put into
agreement with the experimental data using various model
parameterizations for the $U$--matrix. They all  lead to the same
qualitative  predictions, which reflect general properties of this
unitarization scheme.  Originally the Regge-pole model was used to
get an explicit form of $U$--matrix \cite{tuh} and a good
description of the experimental data has been obtained in this
model \cite{edn} as well as in its Dipole-Pomeron modification
\cite{jenk}.

We use the model for the $U$--matrix  based on the ideas of chiral
quark approaches \cite{chpr}. It is in a good agreement with the
data \cite{lhc,pras,univ} and is also applicable to the large
angle scattering. We would like to stress here that the
qualitative conclusions of the present paper do not depend on the
particular $U$--matrix parameterization.

 In this model the picture of a
hadron consisting of constituent quarks embedded
 into quark condensate is used. This picture implies that overlapping
  and interaction of
peripheral clouds   occur at the first stage of hadron interaction.
Nonlinear field couplings  could transform then the kinetic energy to
internal energy and
as a result massive
virtual quarks appear in the overlapping region. These quarks generate an effective
field.
Valence constituent quarks  located in the central part of hadrons are
supposed to scatter simultaneously in a quasi-independent way in this
 field.
Massive virtual quarks play a role of scatterers for the valence
quarks in elastic scattering and their hadronization leads to soft
production process of secondary particles in the central region
\cite{jpg}. The number of such scatterers was estimated
 \begin{equation} \tilde{N}(s,b)\,\propto
\,\frac{(1-\langle k_Q\rangle)\sqrt{s}}{m_Q}D_C(b),
\label{Nsbt}
\end{equation}
under assumption that  part of hadron energy carried by the outer
condensate clouds  released in the overlap region
 to generate massive quarks,
where $m_Q$ is a constituent quark mass and $\langle k_Q\rangle $
-- an average fraction of hadron  energy carried  by  the
constituent valence quarks. Function $D_C(b)$ is a convolution of
the two condensate distributions $D^{h_1}_c({b})$ and
$D^{h_2}_c({b})$ inside the hadron $h_1$ and $h_2$.

We will consider for simplicity the case of a pure imaginary amplitude,
i.e. $U\to iu$.
The
function $u(s,b)$  is represented in the model as a product of the
averaged quark amplitudes $\langle f_Q \rangle$,
\begin{equation} u(s,b) =
\prod^{N}_{i=1} \langle f_{Q_i}(s,b)\rangle \end{equation} in
accordance  with assumed quasi-independent  nature  of the valence
quark scattering, $N$ is the total number of valence quarks in the
colliding hadrons.  The essential point here
is the rise with energy of the number of the scatterers  like
$\sqrt{s}$. The $b$--dependence of the function
$\langle f_Q \rangle$  has a simple form
$\langle f_Q(b)\rangle\propto\exp(-{m_Qb}/{\xi} )$.
The generalized
reaction matrix  gets
the following  form
\begin{equation} u(s,b) = g\left (1+\alpha
\frac{\sqrt{s}}{m_Q}\right)^N \exp(-\frac{Mb}{\xi} ), \label{x}
\end{equation} where $M =\sum^N_{Q=1}m_Q$.
Here $m_Q$ is the mass of constituent quark, which is taken to be
$0.35$ $GeV$\footnote{Other parameters have the following values:
 $g=0.24$, $\xi=2.5$, $\alpha=0.56\cdot 10^{-4}$ which have been obtained from the fit to
  the total
 $hp$ cross sections \cite{pras}.}.

 This model provides  linear
dependence on $\sqrt{s}$ for the total cross--sections,
 i.e.
 $\sigma_{tot}=a+c\sqrt{s}$
    in the limited energy
range $\sqrt{s}\leq 0.5$ TeV. Such behaviour
 and model predictions
for higher energies
 are as it was already mentioned in
 agreement (Fig. 1) with the experimental data on total, elastic and
 diffractive scattering cross-sections \cite{pras,univ}.
\begin{figure}[h]
\begin{center}
  \resizebox{6cm}{!}{\includegraphics*{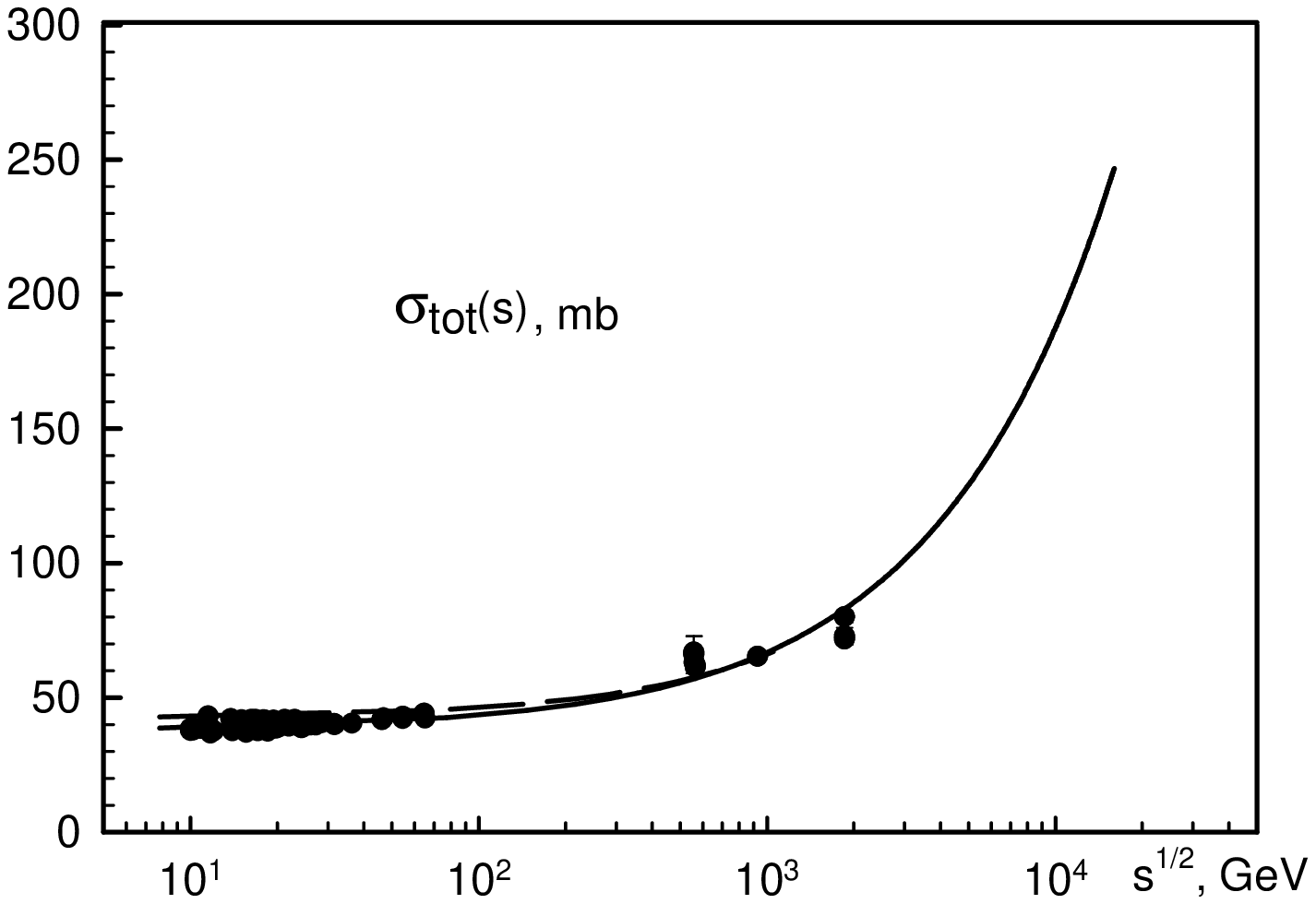}}\;\;\quad
  \resizebox{6cm}{!}{\includegraphics*{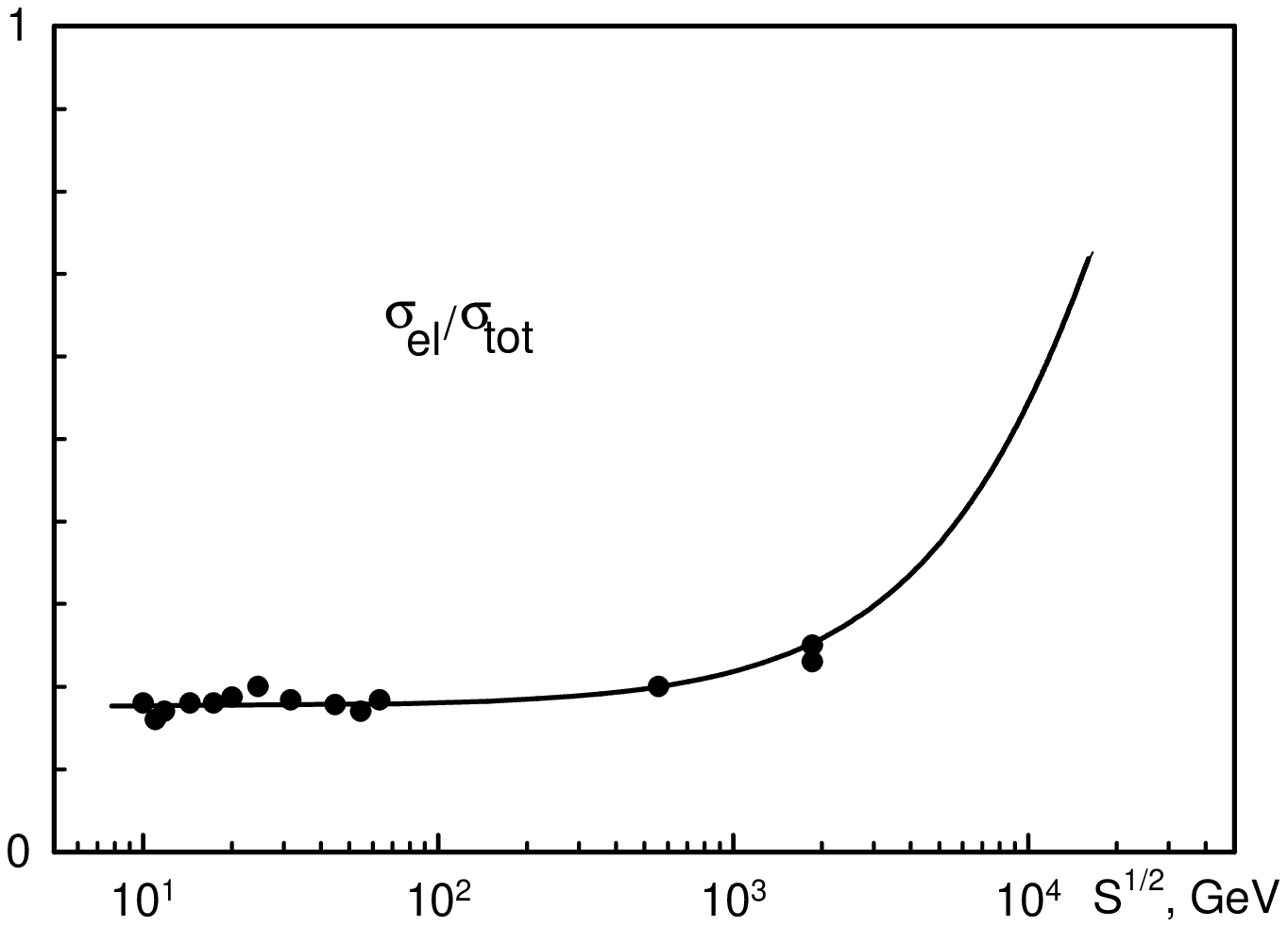}}
\end{center}
\caption{Total and ratio of elastic to total cross-sections of $pp$ and
$\bar p p$--interactions}
\label{ts}
\end{figure}

 This unitarization approach leads to the following asymptotical dependencies
 $\sigma_{tot}\propto \ln^2 s$ and $\sigma_{inel}\propto \ln s$, which are
 the same for the various  models and  reflect essential properties
 of this unitarization scheme.

Thus, now the probability $P(s,b)=|S(s,b)|^2$ can be calculated in a
straightforward way, i.e we use for the function $u(s,b)$ formula (\ref{x})
with parameter fixed from the total cross section fit and relation of
$S(s,b)$ and $U$-matrix (\ref{prob}). The impact parameter dependence of $P(s,b)$
for the different energies is presented on Fig. 2.
\begin{figure}[h]
\begin{center}
\includegraphics*[scale=0.6]{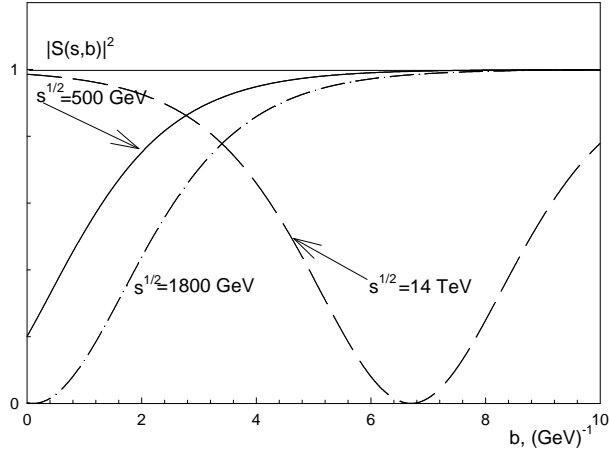}
 \end{center}
\caption{\small\it Impact parameter dependence of probability $P(s,b)=|S(s,b)|^2$
 at three values of energy $\sqrt{s}=500$ GeV (shadow scattering mode),
 $\sqrt{s}=1800$ GeV (black disk limit at $b=0$) and
$\sqrt{s}=14$ TeV (shadow and antishadow scattering modes).}
 \end{figure}

To calculate the gap survival probability $\langle |S|^2\rangle$
we need to know probability of hard interactions $D_H(b)$.
To be specific we  consider the hard central production
processes
\begin{equation}\label{dpe}
p+p\to p+gap+(Higgs\;\mbox{or}\;jj)+gap+p
\end{equation}
The interest in such processes is related to the clear experimental signature and
significant signal-to-background ratio.

We can write down
 the probability of hard interactions in the model as a convolution
\begin{equation}
D_H({b})=\sigma_H\int D_c^{h_1}({\bf b}_1)w_H({\bf b}+{\bf b}_1-{\bf b}_2)D_c^{h_2}({\bf b}_2)
d{\bf b}_1 d {\bf b}_2,
\end{equation}
where   $w_H({\bf b}+{\bf b}_1-{\bf b}_2)$ is the probability of
hard condensate (parton) interactions. It is natural to assume that the probability
$w_H$ has much
 steeper impact parameter dependence than the functions $D_c^{h_i}({ b})$ have and,
 therefore,
 the impact parameter dependence of $w_H$ determines behaviour of $D_H({b})$.
 Thus, we assume a simple exponential dependence for the function $D_H({b})$, i.e.
\begin{equation}\label{dh}
D_H({b})\simeq \sigma_H\exp(-M_Hb),
\end{equation}
where mass $M_H$ is determined by  the hard scale of the process.
We perform numerical calculations of
the gap survival probability $\langle |S|^2\rangle$ using Eqs.
(\ref{gap}), (\ref{prob}) and (\ref{dh}).

We take the hard scale
$M_H$ to be determined by the mass of $J/\Psi$-meson, i.e.
$M_H=M_{J/\Psi}$. This choice is in accord with the fact
 that production of $J/\Psi$-meson can be treated
as a hard process and therefore its mass set a hard scale
\cite{bartels}. Lower values of $M_H$ lie in the soft region
\footnote{In Eq. (\ref{dh}) $\sigma_H$ can be interpreted as a probability of
the hot spot formation under condensate interaction and
$R_H\simeq1/M_H$ is the radius of this hot spot.}.
It should be noted that
numerical results are rather stable and depend weakly on the scale $M_H$, when it is
in the hard region, i.e. $M_H\geq M_{J/\Psi}$
  e.g. for illustration we used the value
$M_H=8$ GeV and it leads to slightly lower values of the gap survival
probability at low energies.
Results of calculations are presented
in Fig. 3.
\begin{figure}[h]
\begin{center}
\includegraphics*[scale=0.6]{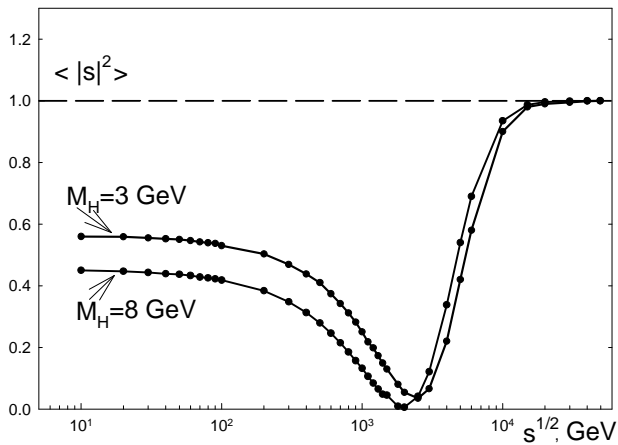} \vspace*{-0.5cm}
 \end{center}
\caption{\small\it Energy dependence of gap survival probability $\langle |S|^2\rangle$}
 \end{figure}
One can notice  that the gap survival probability reaches its
minimal values at the Tevatron highest energy. It is not
surprising since the scattering at this energy is very close to
the black disk limit.

Asymptotical behaviour of the gap survival probability has the form:
\begin{equation}\label{asym}
\langle |S|^2\rangle\simeq 1-\frac{\xi M_H}{2m_Q}s^{-\frac{\xi M_H}{2m_Q}}\ln s.
\end{equation}
The two unitarization schemes ($U$--matrix and eikonal)
 lead to
 different predictions for the gap survival probability
in the limit $s\to\infty$; eikonal unitarization predicts $\langle
|S|^2\rangle=0$ at $s\to\infty$, while $U$--matrix formalism gives
$\langle |S|^2\rangle=1$. Latter is a result of transition to the
antishadow scattering mode in the $U$--matrix unitarization
\cite{bds}, when the amplitude becomes $|f(s,b)|> 1/2$   (in the
case of imaginary eikonal the scattering amplitude never
exceeds the black disk limit
 $|f(s,b)|\leq 1/2$). It should be noted that the Froissart-Martin bound
 implies unitarity (not black disk) limit for the partial amplitudes.
When the amplitude exceeds the black disk limit (in central
collisions at high energies) then the scattering at such impact
parameters turns out to be of an  antishadow nature.
 In this antishadow scattering mode
 the elastic amplitude increases with decrease of the inelastic
 channels contribution at small impact parameters and most of
 inelastic interactions occur in the
 peripheral region; the inelastic overlap function  has a peripheral
 impact parameter profile, which is a main reason of the large gap
 survival probability.

 Numerical predictions for the gap survival probability
 obtained here   depend on the
  particular parameterization for $U$--matrix,
 but the qualitative picture of the
 energy behaviour  of $\langle |S|^2\rangle$ reflects transition to the new
 scattering mode at the LHC energies and is valid for various
   $U$--matrix parameterizations which provide rising total cross--sections.
  One should note that the numerical values of
 $\langle |S|^2\rangle$ at the Tevatron energies are in a qualitative agreement
 with a number of  quantitative calculations performed in the eikonal
 approaches (cf. \cite{khoze}).

  In the sense  of gap survival probability
  situation should be more favorable at the LHC energies
 since the obtained numerical values of $\langle |S|^2\rangle$ at these
  energies are close to unity and this should lead to much higher cross--sections
 (by factor of $40$ compare to the calculations based on the gap survival probability
 estimations in the framework of the eikonal model \cite{martin} )
  e.g. for Higgs production in double diffractive processes compared to the
   values obtained with eikonal based estimations of the gap survival probability.

Thus, the antishadowing  which appearance  is expected
at the LHC energies, should be correlated with enhancement of Higgs production
cross--section in double diffractive scattering and this would
significantly help in detecting of Higgs boson.

\end{document}